\documentclass[a4paper, 11pt]{article}

\usepackage[top=1.25in, bottom=1.25in, left=1in, right=1in]{geometry}
\usepackage[affil-it]{authblk}
\usepackage{mathtools}
\usepackage{float}
\usepackage{amsmath}
\usepackage[title]{appendix}
\DeclareMathOperator*{\argmin}{argmin}
\DeclareMathOperator*{\argmax}{argmax}
\usepackage{graphicx}
\usepackage{multirow}
\usepackage{mathtools}
\usepackage{multicol}
\usepackage{amsfonts}
\usepackage{nccmath}
\usepackage{amsmath}
\usepackage{amssymb, amsthm, mathrsfs}
\usepackage[T1]{fontenc}
\usepackage{enumerate}
\usepackage{caption}
\usepackage{natbib}
\usepackage{subcaption}
\usepackage[utf8]{inputenc}
\usepackage[english]{babel}

\newtheorem{theorem}{Theorem}

\usepackage{hyperref}
\hypersetup{citecolor=blue,colorlinks=true}
\usepackage{autonum}

\newcounter{example}[section]
\newenvironment{example}[1][]{\refstepcounter{example}\par\medskip
	\noindent \textbf{Example~\theexample. #1} \rmfamily}{\medskip}

\newtheorem*{remark}{Remark}

\providecommand{\keywords}[1]{\textbf{\textit{Keywords---}} #1}
\begin{document}
	%\begin{frontmatter}
	
	\title{Robust Inference Using the Exponential-Polynomial Divergence }
	
	\author[1]{Pushpinder Singh}\author[2]{Abhijit Mandal} \author[1]{Ayanendranath Basu}
	\affil[1]{Interdisciplinary Statistical Research Unit, Indian Statistical Institute, Kolkata, India} 
	\affil[2]{Department of Mathematical Sciences, University of Texas at El Paso, El Paso, USA}
	%\date{\today}
	\maketitle
	
	%\end{frontmatter}
	\begin{abstract}
		Density-based minimum divergence procedures represent popular techniques in parametric statistical inference. They combine strong robustness properties with high (sometimes full) asymptotic efficiency. Among density-based minimum distance procedures, the methods based on the Br{\`e}gman-divergence have the attractive property that the empirical formulation of the divergence does not require the use of any non-parametric smoothing technique such as kernel density estimation. The methods based on the density power divergence (DPD) represent the current standard in this area of research. In this paper we will present a more generalized divergence which subsumes the DPD as a special case, and produces several new options providing better compromises between robustness and efficiency. 
	\end{abstract}
	
	\keywords{Br{\`e}gman divergence, Density power divergence, M-estimator, Robustness.}

	\section{Introduction}
	In parametric statistical inference, the likelihood based methods have several asymptotic optimality properties. Under model misspecifications, or under the presence of outliers, all classical procedures including maximum likelihood may, however,  be severely affected and lead to a distorted view of the true state of nature. For the big data scenario of the present times, a certain amount of noise is never unexpected, but even a small amount of it may be sufficient to severely degrade the performance of the classical procedures.
	
	Another approach to parametric estimation is based on minimizing density-based distances between the data density and the proposed model density; this approach  generally combines high efficiency (sometimes full asymptotic efficiency) with strong robustness properties. Interestingly, the maximum likelihood estimator (MLE) itself also belongs to this class of density-based minimum distance estimators, being the minimizer of a version of the Kullback-Leibler divergence  \citep{kullback1951}. Another such divergence is the Hellinger distance and the estimator generated by minimizing this divergence is both highly robust (compared to the MLE) and  is also first-order efficient \citep{beran1977}. However the minimum  Hellinger distance estimation method (and similar estimation methods based on $\phi$-divergences) is burdened by the fact that a non-parametric smoothing technique is inevitably necessary for the construction of an estimate of the data density for this procedure under continuous models. Apart from computational difficulties and the tricky bandwidth selection issue, the slow convergence of the non-parametric density estimate to the true density in high dimensions poses a major theoretical difficulty.

	An alternative density-based minimum distance procedure is presented by the class of Br{\`e}gman divergences \citep{Bregman1967200}. This divergence class is characterized by a strictly convex function, and does not need any non-parametric smoothing for its empirical construction; see \cite{jana2019characterization}. 
	Although the estimators obtained by the minimization of Br{\`e}gman divergences are generally not fully efficient, they often combine high asymptotic efficiency with a strong degree of robustness and outlier stability.
	Most minimum Br{\`e}gman divergence estimators also have bounded influence function, a property which is not shared by the minimum $\phi$-divergence estimators. See \cite{basu2011statistical} and \cite{pardo2006statistical} for more details on robust parametric inference based on $\phi$-divergences (or disparities).
	
	The DPD family \citep{basu1998robust}  is a prominent subclass of Br{\`e}gman divergences, and has had a significant impact on density-based minimum distance inference in recent times. This divergence family is defined by a class of convex functions indexed by a non-negative tuning parameter $\gamma$. Larger values of $\gamma$ lead to divergences which endow the corresponding minimum-divergence estimator with greater outlier stability. In the following we will refer to the minimum density power divergence estimator as the MDPDE, and tag on the $\gamma$ symbol wherever necessary. Our aim is to improve upon the MDPDE($\gamma$) in terms of robustness and efficiency. It is useful to note that the minimum Br{\`e}gman divergence estimator (and hence the MDPDE) belongs to the class of M-estimators as defined in, for example, \cite{hampel2011robust} or \citet{maronna2019robust}. As a consequence, the asymptotic properties of the estimators based on the newly defined divergences may be obtained from the well established M-estimation literature. The minimum density power divergence estimation method for independent and identically distributed (IID) data have been discussed in \citet{basu1998robust,basu2011statistical} and that for independent non-homogeneous (INH) data in \citet*{ghosh2013robust}. We will perform similar exercises for our newly defined divergence family, and show that the resulting procedures can become useful robust tools for the applied scientist. 
	
	\section{The Br{\`e}gman Divergence and Related Inference} \label{sec:Breg_div}
	The Br{\`e}gman divergence was originally proposed as a measure to define a distance
	between two points in $\mathbb{R}^{d}$. It is a divergence measure,  but not a metric in the true sense of the term, as it generally does not satisfy  the triangle inequality and may not even be symmetric in its arguments. Let $B\colon \mathbb{R}^{d} \rightarrow \mathbb{R} $ be a twice continuously differentiable, strictly convex function defined on a closed convex set in $\mathbb{R}^{d} $.
	The Br{\`e}gman divergence associated with the strictly convex function $B$  for $p,q \in \mathbb{R}^{d} $ is defined as
	\begin{equation}\label{eqn:General Bregman Form}
	D_{B}(p,q) = B(p) - B(q) - \langle B'(q),p-q \rangle, 
	\end{equation}
	where $B'$ represents the derivative of $B$ with respect to its argument. For two given density functions $g$ and $f$, the Br{\`e}gman divergence between them is defined as
	\begin{equation} \label{eqn:Bregman density form}
	D_{B}(g,f) = \int_{x} \Big\{ B(g(x)) - B(f(x)) - (g(x)-f(x))B'(f(x)) \Big\} dx.
	\end{equation}
	The function $B$, in the above case, is clearly not uniquely defined due to the linearity property of the integral, as both $B(y)$ and  $B(y) + ay + b$ give rise to the exact same  divergence for any real constants $a$ and $b$. Here we explore the general estimation procedure to  find the minimum Br{\`e}gman divergence estimator  for  any convex $B$ function. Assume that an IID random sample $X_1, X_2, \ldots , X_n $ is available from the true distribution $G$, and we try to model this distribution by a parametric family $\mathscr{F}=\{F_{\theta}: \theta \in \Theta \subset {\mathbb R}^p \} $ where $\theta$ is unknown but the functional form of $F_{\theta}$ is known to us. In such a scenario, the estimation of the parameter $\theta$ consists in  choosing the model density  $f_{\theta} $ which is closest to the  data density in the minimum Br{\`e}gman divergence sense. Let $g$ and $f_{\theta}$ be the probability densities  of $G$ and $F_{\theta}$ respectively. Then the Br{\`e}gman divergence between $g$ and $f_{\theta}$ will be as given in Equation (\ref{eqn:Bregman density form}) with $f$ replaced by $f_\theta$.

	We wish to use the minimum Br{\`e}gman divergence approach for the estimation of the unknown parameter $\theta$. Notice that we cannot directly obtain the Br{\`e}gman divergence between $g$ and $f_\theta$ for the purpose of this minimization, as the density $g$ is unknown. So we need an empirical estimate of this divergence, which can then be minimized over $\theta \in \Theta$.  After discarding the terms of the above divergence (objective function) that are independent of $\theta$, the only term that needs to be empirically estimated is $\int B'(f_\theta(x)) g(x) dx$, which can be estimated by the corresponding sample mean $\frac{1}{n} \sum_{i=1}^n B'(f_\theta(X_i))$, so the empirical objective function for the minimization of $D_B(g, f_\theta)$ is now given by 
	\begin{equation}
	\int_x \big\{ B'(f_\theta(x))f_\theta(x) - B(f_\theta(x)) \big\} dx - \frac{1}{n} \sum_{i=1}^n B'(f_\theta(X_i)).  
	\label{gen_Bre_est}
	\end{equation}
	Let $u_{\theta}(x) = \nabla_{\theta} \log(f_{\theta}(x))$ be the likelihood score function of the model being considered where $\nabla_{\theta}$ represents the gradient with respect to $\theta$. Under appropriate differentiability conditions, the minimizer of this empirical divergence over $\theta \in \Theta$ is obtained as a solution to the estimating equation
	
	\begin{equation}\
	\frac{1}{n} \sum_ {i=1}^{n} u_{\theta}(X_{i})B''(f_{\theta}(X_{i}))f_{\theta}(X_{i})- \int_{x}^{} u_{\theta}(x)B''(f_{\theta}(x))f_{\theta}^{2}(x) dx = 0.
	\end{equation}
	This may be viewed as being in the general weighted likelihood equation form given by 
	\begin{equation}
	\frac{1}{n} \sum_ {i=1}^{n} u_{\theta}(X_{i})w_{\theta}(X_{i}) - \int_{x}^{} u_{\theta}(x)w_{\theta}(x)f_{\theta}(x) dx = 0,
	\label{wt_lik}
	\end{equation}
	where the relation between the Br{\`e}gman function $B$ and the weight function $w_{\theta}$ is given as
	\begin{equation}
	w_{\theta}(x) = w(f_\theta(x)) = B''(f_{\theta}(x))f_{\theta}(x) .
	\label{wx}
	\end{equation}
	The non-negativity of the above weight function is secured by the convexity of the $B$ function with the non-negativity of the density function. Some of the major density-based divergences that can be obtained from the Br{\`e}gman divergence using different $B$ functions are the following.
	\begin{enumerate}
		\item   $B(y) = y\log(y)-y$: This generates the \textit{Kullback-Leibler} divergence given by 
		\begin{equation}
		D_{KL}(g,f_{\theta})= \int_{x}^{} g(x) \log \bigg( \frac{g(x)}{f_{\theta}(x)}  \bigg) dx.
		\end{equation}
		Under our parametric setup, its estimating equation and  weight function are, respectively,
		\begin{equation}
		\frac{1}{n}\sum_{i=1}^{n} u_{\theta}(X_{i})-\int_{x}^{} u_{\theta}(x)f_{\theta}(x)  dx = 0, \ \ 
		w_\theta(x) = w(f_\theta(x)) = 1.
		\label{est_kl}
		\end{equation}
		
		\item $B(y) = y^2$: This leads to the squared $L_2$ distance 
		\begin{equation}
		L_{2}(g,f_{\theta}) = \int_{x}^{} \big[g(x)-f_{\theta}(x)\big]^{2} dx,
		\end{equation}
		generating, respectively, estimating equation and weight function as
		\begin{equation}
		\frac{1}{n}\sum_ {i=1}^{n} u_{\theta}(X_{i})f_{\theta}(X_{i}) = \int_{x}^{}u_{\theta}(x)f_{\theta}^{2}(x) dx,
		\ \ w_\theta(x) = w(f_\theta(x)) = f_\theta(x).   \label{est_l2}
		\end{equation}
		
		\item $B(y)= (y^{1+\gamma}-1)/\gamma$: This generates the DPD($\gamma$) family given by
		\begin{equation}
		d_{\gamma}(g,f_{\theta}) = \int_x \Big\{ f_{\theta}^{1+\gamma}(x)-\bigg(1 + \frac{1}{\gamma}\bigg)g(x)f_{\theta}^{\gamma}(x)+\frac{1}{\gamma} g^{1+\gamma}(x)\Big\} dx.
		\label{dpd}
		\end{equation}
		In this case its estimating equation and  weight function are given by 
		\begin{equation}
		\frac{1}{n}\sum_{i=1}^{n} u_{\theta}(X_{i})f_{\theta}^{\gamma}(X_{i})-\int_{x}^{} u_{\theta}(x)f_{\theta}^{1+\gamma}(x) dx = 0,
		\ \ w_\theta(x) = w(f_\theta(x)) = f_\theta^\gamma(x).
		\label{est_dpd}
		\end{equation}
		
	\end{enumerate}
	
	It may be noted that the estimating equations (\ref{est_kl}), (\ref{est_l2}) and (\ref{est_dpd}) are all unbiased under the model and have the same general structure as given in Equation (\ref{wt_lik}). The equations differ only in the form of the weight function $w_\theta(x)$. And it is this weight function which determines to what extent the estimating equation is able to control the contribution of the score to the equation. In Equation (\ref{est_kl}) the weight function is identically 1, so that the equation has no downweighting effect over  the score functions of anomalous observations. The $L_2$ case in Equation (\ref{est_l2}), on the other hand, provides a strong downweighting effect by attaching the density function as the weight. The DPD covers a middle ground, by generating a weight of $f^\gamma_\theta(x)$, which produces a smoother downweighting compared to  the $L_2$ case for $\gamma \in (0, 1)$. 
	
	\section{The Exponential-Polynomial Divergence}
	Here our aim is to find a suitable convex function so that we can propose a generalized class of Br{\`e}gman divergences that generates the DPD class as a special case. For this purpose we consider a sophisticated convex function $B$ having the general form
	
	\begin{equation}
	\begin{aligned}
	\label{B}
	B(x) = \beta\frac{(\exp(\alpha x)-1-\alpha x)}{\alpha^2}  + (1-\beta)\frac{{(x^{\gamma+1}-x)}}{\gamma} ,
	\end{aligned}
	\end{equation}      
	where $\alpha $, $\beta$ and $\gamma $ are  the tuning parameters for the system. The function in Equation (\ref{B}) is considered to be a generalization of the generating function for DPD given in Equation (\ref{dpd}). Clearly we recover the DPD with parameter $\gamma$ for  $\beta = 0$, but for non-zero $\beta$ we get a combination of the generating function for the Br{\`e}gman exponential-divergence (BED) \citep{Bexp} and the density power divergence. At $\beta = 1$, we get the BED with tuning parameter $\alpha$. While the value $\beta$ moderates the level of presence (or absence, when $\beta = 0$) of the BED component, $\alpha$ and $\gamma$ represent the BED and the DPD tuning parameters respectively. Note that when $\beta = 0$ and $\gamma \rightarrow 0$, the divergence converges to the Kullback-Leibler divergence. We refer to the divergence produced by Equation (\ref{B}) as the exponential-polynomial divergence (EPD) and we will be using the notation 
	$D_{EP}(g, f_\theta)$ to refer to the exponential-polynomial divergence between the densities $g$ and $f_\theta$. The tuning parameters of these families lie in the regions $\alpha \in {\mathbb R}$, $\beta \in [0,1]$ and $\gamma \geq 0$. In the spirit of the notation employed so far, the $B$-function of the EPD may be seen to be a convex combination of the BED($\alpha$) and DPD($\gamma$) $B$-functions. 
	
	\subsection{Minimum EPD  Estimation as M-Estimation} \label{sec:m_est}
	Consider the parametric setup of Section \ref{sec:Breg_div} and the empirical objective function of the Br{\`e}gman divergence given in Equation (\ref{gen_Bre_est}). Note that, in case of the EPD, this objective function may be written as $\frac{1}{n} \sum_{i=1}^n V_\theta(X_i)$, where 
	\begin{equation}
	\begin{aligned}
	V_{\theta}(x) & = - \frac{\beta}{\alpha} \Big (\exp(\alpha f_\theta(x))-1\Big) - \frac{1-\beta}{\gamma}\Big ( (\gamma+1) f_\theta ^\gamma (x) - 1 \Big)\\
	&+ \int_t \bigg[ \frac{\beta}{\alpha^2} \{  \exp(\alpha f_\theta(t)) \big (\alpha f_\theta (t) -1\big ) + 1  \} + (1-\beta) f_\theta^{\gamma+1} (t)     \bigg]
	dt.
	\end{aligned}
	\label{V_theta}
	\end{equation}
	As $X_1, X_2,  \cdots, X_n$ are independent and identically distributed observations, $V_\theta(X_i)$, $i =1,2, \cdots, n$ are independent and identically distributed as well. Under differentiability of the model the estimating equation is
	\begin{equation}
	\sum_{i=1}^{n}\psi(X_{i},\theta)=0.
	\label{est_eqn}
	\end{equation}
	Direct calculations show that for the EPD the  associated $\psi$ function has the form \begin{equation}
	\psi(x, \theta) = T_\theta(x) - E_{f_\theta}(T_\theta(X)), 
	\end{equation}
	where 
	\begin{equation}
	T_\theta(x) = u_\theta(x) \Big\{ \beta f_\theta(x) \exp(\alpha f_\theta(x)) 
	+ (1-\beta)(\gamma+1) f_\theta^\gamma(x) \Big\}. 
	\label{tx}
	\end{equation}
	In particular, for a location model, the estimating equation reduces to $\sum_{i=1}^n T_\theta(X_i) = 0$. 
	%\begin{equation}
	%\sum_{i=1}^n T_\theta(X_i) = 
	%\sum_{i=1}^n u_\theta(X_i) \{ \beta f_\theta(X_i) \exp(\alpha f_\theta(X_i)) 
	%+ (\gamma+1) f_\theta^\gamma(X_i) \} = 0. 
	%\end{equation}
	The above description shows that the minimum EPD estimator (MEPDE) is an M-estimator (which is indeed true for all minimum Br{\`e}gman divergence estimators). The functional  $T_{(\alpha,\beta,\gamma)}(G) $, defined through the relation $T_{(\alpha, \beta, \gamma)}(G) = \argmin_{\theta \in \Theta } D_{EP}(g, f_\theta)$,  is easily seen to be Fisher consistent, so that,  $T_{(\alpha,\beta,\gamma)}(F_{\theta})= \theta$. If the distribution $G$ is not in the parametric family $\mathscr{F}$, then  $T_{(\alpha,\beta,\gamma)}(G) $ is the solution of the equation 
	\begin{equation}
	E_g(T_\theta(X)) = E_{f_\theta}(T_\theta(X)). 
	\end{equation}
	In this case we will refer to this solution as the best fitting parameter and denote it by $\theta^g$.

	\subsection{Asymptotic Properties}
	We define the empirical objective function to be 
	\begin{align}
	H_{n}(\theta)=  n^{-1} \sum_{i=1}^{n} V_{\theta}(X_{i}),
	\end{align}
	where $V_\theta (x)$ is as defined in Equation (\ref{V_theta}).
	The theoretical analogue of $H_n(\theta)$ is given by  
	\begin{align}
	\begin{split}
	H(\theta) &= - \int_x \bigg[ \frac{\beta}{\alpha} \big (\exp(\alpha f_\theta(x))-1\big) dx + \frac{1-\beta}{\gamma}\big ( (\gamma+1) f_{\theta} ^\gamma (x) - 1 \big) \bigg] g(x) dx \\ \nonumber
	& + \int_x \bigg[ \frac{\beta}{\alpha^2} \Big\{   \exp(\alpha f_{\theta}(x)) \big (\alpha f_\theta (x) -1\big ) + 1  \Big\} + (1-\beta)f_\theta^{\gamma+1} (x)     \bigg]   
	dx.
	\end{split}  \nonumber
	\end{align}
	We define the information function of the model as  $ i_{\theta}(x)=-\nabla u_{\theta}(x)$, and further define the quantities $K(\theta)$, $\xi(\theta)$ and $J(\theta)$ as
	
	\begin{equation}
	\begin{aligned}
	K(\theta) &= \int_x u_{\theta}(x)u_{\theta}^{T}(x)\Big\{\beta f_{\theta}(x)\exp(\alpha f_{\theta}(x)) + (1-\beta)(\gamma+1)f_{\theta}^{\gamma}(x)\Big\}^{2}g(x) dx - \xi(\theta)\xi^{T}(\theta) ,    \\
	\xi(\theta) &= \int_x u_{\theta}(x)\Big\{\beta f_{\theta}(x)\exp(\alpha f_{\theta}(x)) + (1-\beta)(\gamma+1)f_{\theta}^{\gamma}(x)\Big\} g(x) dx, \\
	J(\theta) & =  \beta \int_x f_{\theta}^{2}(x) \exp(\alpha f_{\theta}(x)) u_{\theta} (x) u_\theta ^{T} (x) dx + (1-\beta)(\gamma+1) \int_x f_{\theta}^{\gamma+1}(x) u_{\theta}(x) u_\theta ^T (x) dx\\ 
	&   + (1-\beta)(\gamma+1)\int_x \big( g(x)-f_{\theta}(x)\big) {\Big\{ i_{\theta}(x)-\gamma u_{\theta}(x) u_\theta ^T (x)  \Big\} f_{\theta}^{\gamma}(x)} dx\\
	&  +\beta \int_x (g(x)-f_{\theta}(x))\Big\{ i_{\theta}(x) - u_{\theta}(x) u_\theta^T (x)  \Big\}  f_{\theta}(x) \exp(\alpha f_{\theta}(x)) dx \\
	&  - \alpha \beta \int_x \big( g(x)-f_{\theta}(x)\big) f_{\theta}^{2}(x) \exp(\alpha f_{\theta}(x)) u_{\theta}(x) u_\theta^T (x) dx.
	\end{aligned}
	\label{J_K_psi}
	\end{equation}
	\begin{theorem}\label{theorem1}
		Under the  conditions (A1)--(A5) given in  Appendix A
		\begin{enumerate}
			\item[(a)] 
			The MEPDE estimating equation given by (\ref{est_eqn}) has a consistent sequence of roots of $ \hat{\theta}_{n}. $ %$\hat{ \theta} =\hat{\theta}_{n}. $
			
			\item[(b)] $\sqrt{n}(\hat{\theta}_{n}-\theta^{g}) $ has an asymptotic multivariate normal distribution with mean (vector) zero and covariance matrix  $J^{-1}KJ^{-1}$ where $J$ and $K$ are defined in Equation (\ref{J_K_psi}), and evaluated at $\theta = \theta^g$.
		\end{enumerate} 
	\end{theorem}
	
	The proof is a relatively straightforward extension of Theorem 6.4.1 of \cite{lehmann1983theory}, and is omitted. The result can also be obtained, as indicated, from the M-estimation approach, but the conditions of this proof are slightly weaker.

	\subsection{Influence Function, Gross Error Sensitivity and Asymptotic Efficiency}
	An useful advantage of the representation of the minimum Br{\`e}gman divergence estimator as an M-estimator is the straightforward computation of its influence function. Another important measure of robustness,  available from the influence function  is the gross error sensitivity (GES). Based on the nature of its influence function or GES, we can comment on the robustness properties of the associated MEPDE. Simple calculations show that the influence function of the MEPD functional $T_{(\alpha,\beta,\gamma)}(\cdot)$ has the form
	\begin{equation} 
	IF(y,T_{(\alpha,\beta,\gamma)},G)= J^{-1} \big[u_{\theta}(y)\big(\beta f_{\theta}(y) \exp(\alpha f_{\theta}(y)) + (1-\beta)(\gamma+1)f_{\theta}^{\gamma}(y)\big)-\xi\big],
	\end{equation}
	where $\xi$ and $J$,  as in Equation (\ref{J_K_psi}), are evaluated at $\theta = \theta^g$. Under the assumption that $J$ and $\xi$ are finite, this influence function is bounded only if the quantity $\big\{u_{\theta}(y)\big(\beta f_{\theta}(y) \exp(\alpha f_{\theta}(y)) + (1-\beta)(\gamma+1)f_{\theta}^{\gamma}(y)\big)\big\}$ is bounded in $y$. This is indeed true for all standard models for $\beta \in [0, 1]$, $\alpha \in \mathbb{R}$ and $\gamma>0$. The GES of the  functional $ T_{(\alpha, \beta, \gamma)}(G)$ is
	\begin{equation}
	\gamma^{*}( T_{(\alpha, \beta, \gamma)}(G), \theta ) = \argmax_{y}  \{IF(y,T_{(\alpha, \beta, \gamma)},G)\}.
	\end{equation}
	The influence function of the MDPDE is bounded for $\gamma > 0$, and that for the MEPDE is bounded for $\gamma > 0$ and any finite $\alpha$ and $\beta \in [0, 1]$, so that our functional has finite GES for the indicated set of tuning parameters. It should be noted that the influence function and the GES for the MLE are unbounded. 
	
\begin{figure}
		\begin{center}\vspace{-.2in}
			\includegraphics[scale=0.75]{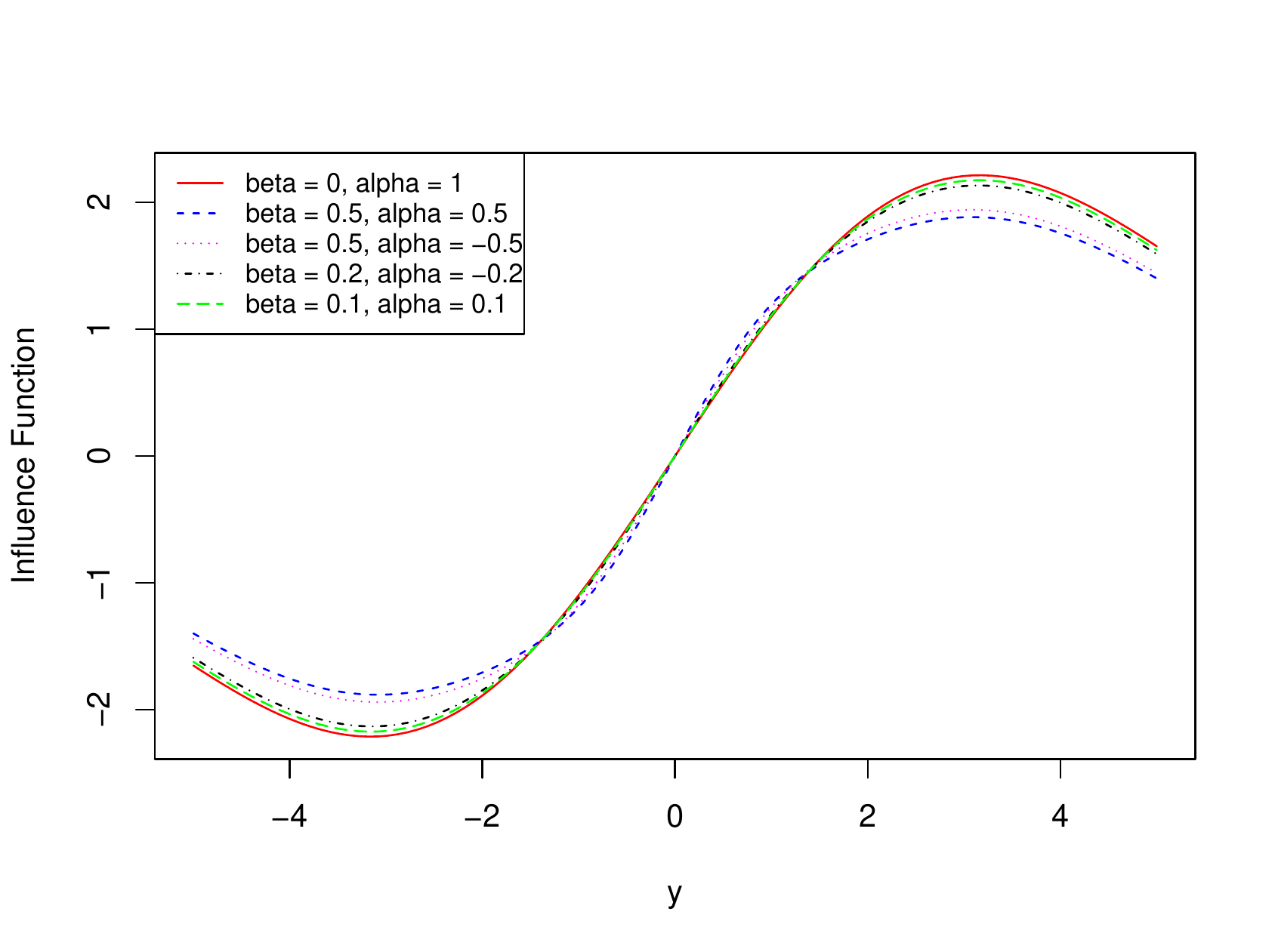}
			\caption{Influence functions of different MEPDEs in the $N(\mu, 1)$ model.}
			\label{fig:inf}
		\end{center}
	\end{figure}

	As an example, we consider a particular case for our illustration with the influence function. In Figure \ref{fig:inf} we present the influence function of the MEPDE functional for the mean of a normal random random variable under the $N(\mu, 1)$ model, where $N(0,1)$ is the true distribution. The value of $\gamma$ is fixed to be 0.1 in this example, and while the choice $\beta = 0$ (irrespective of the value of $\alpha$) refers to MDPDE(0.1), the figure shows that at different choices of $\alpha$ and nonzero $\beta$ at the same value of $\gamma$ ($=0.1$), substantially lower peaks for the influence function (and hence smaller GES values) may be attained for the corresponding MEPDE. While there is no doubt that substantial further investigation will be necessary to get an overall feeling of the stability of the estimator for different choices of the triplet ($\alpha$, $\beta$, $\gamma$), it is clear that other parameter combinations can increase the strength of downweighting, without altering the value of $\gamma$. In Figure \ref{fig:inf} we have refrained from  adding the plot for the unbounded MLE to avoid unnecessary cluttering of the graph. 
	
It may be noted, however,   that the above exercise is not intended to suggest that for each fixed divergence within the DPD family there is a better choice of a divergence within the EPD family (with the same value of $\gamma$) which can dominate the former in terms of all the possible goodness measures. Indeed, detailed numerical calculations show that often there may not be another member of the EPD family which may have improved efficiency compared to the corresponding DPD with the same value of $\gamma$. On the other hand, detailed calculations appear to suggest that given a divergence within the DPD family, often there may be another divergence within the EPD family, not necessarily with the same value of $\gamma$, which might provide better metrics than the former.            
	
	The asymptotic variance of $\sqrt{n}$ times the MEPDE($\alpha,\beta,\gamma$) can be estimated through the influence function using the asymptotic distribution of M-estimators; see, eg., \cite{hampel2011robust}. Let $R_i(\theta)$, $i= 1,\cdots,n$,  be the quantity  $u_{\theta}(X_i)\big\{\beta f_{\theta}(X_i) \exp(\alpha f_{\theta}(X_i)) + (1-\beta)(\gamma+1) f_{\theta}^{\gamma}(X_i)\big\} $ at the data point $X_{i}$ for $\theta = \hat{\theta}_{(\alpha,\beta,\gamma)}$, the estimator minimizing the divergence. We can estimate the $J$ matrix by $J(G_n)$ obtained by substituting $G$ with $G_n$, the empirical distribution function, in the expression of $J$. Then, a consistent estimate of asymptotic variance of the MEPDE may be obtained as
	\vspace{0.4cm}$J^{-1}(G_n)\big\{(n-1)^{-1}\sum_{i=1}^{n}R_{i} R_{i}^{T}\big\} J^{-1}(G_n)$. 
	
	\subsection{The Weight Function}
	
\begin{figure}
		\centering%
		\begin{tabular}{cc}
			\includegraphics[scale=0.28]{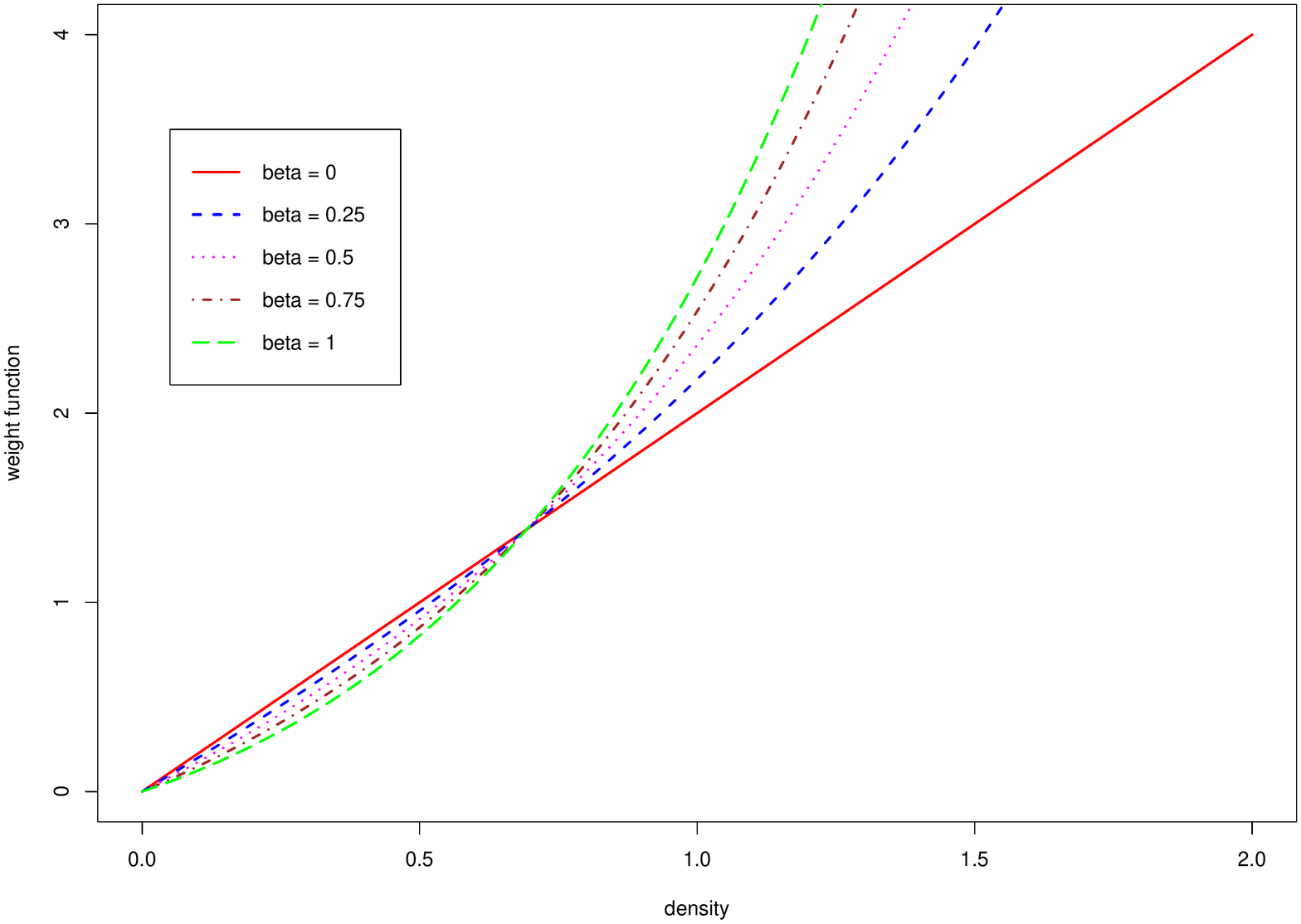} & \includegraphics[scale=0.28]{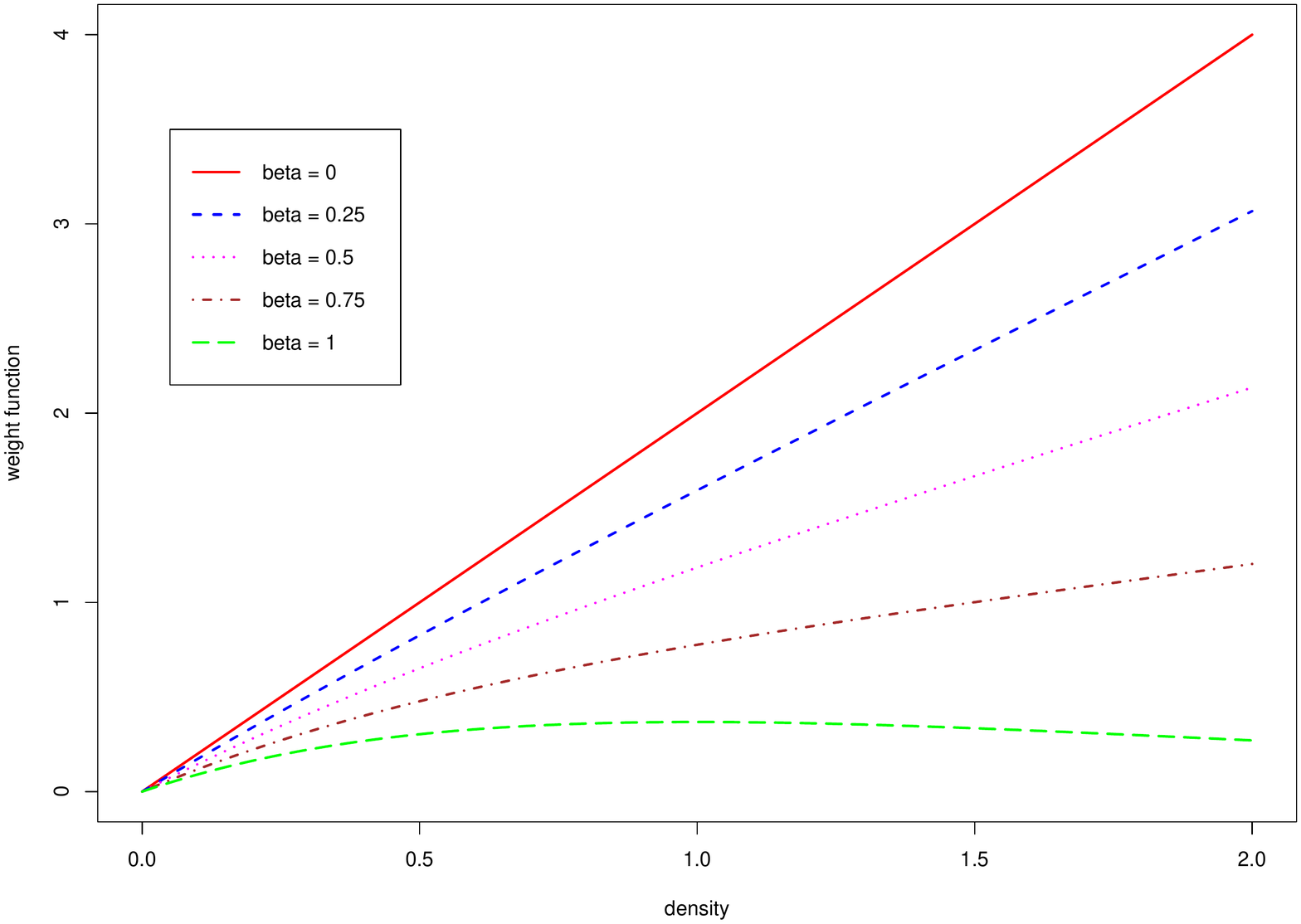}\vspace{-.2cm}\\ \vspace{-.1cm}
			(a)  & (b)\\
			\includegraphics[scale=0.28]{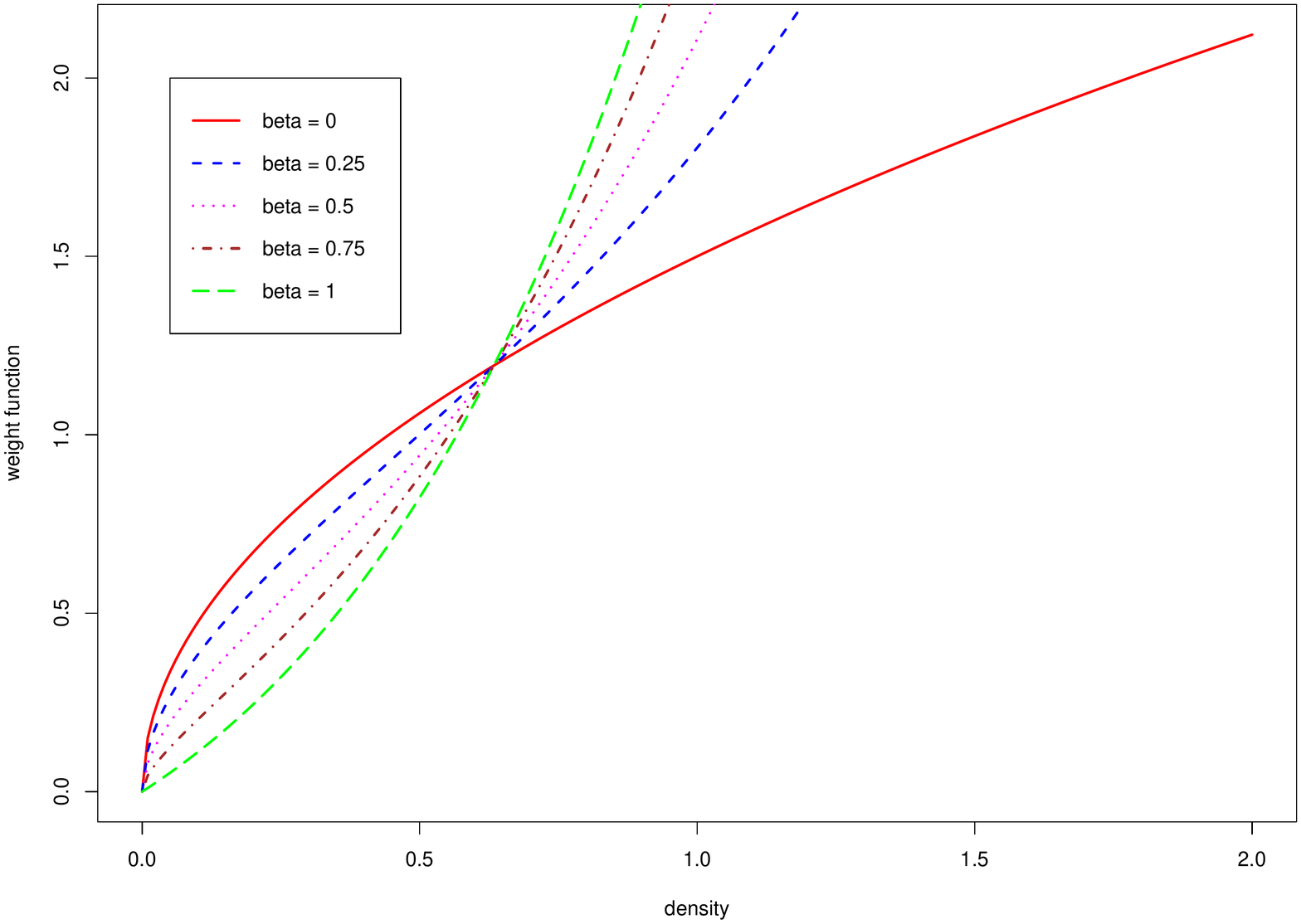} & \includegraphics[scale=0.28]{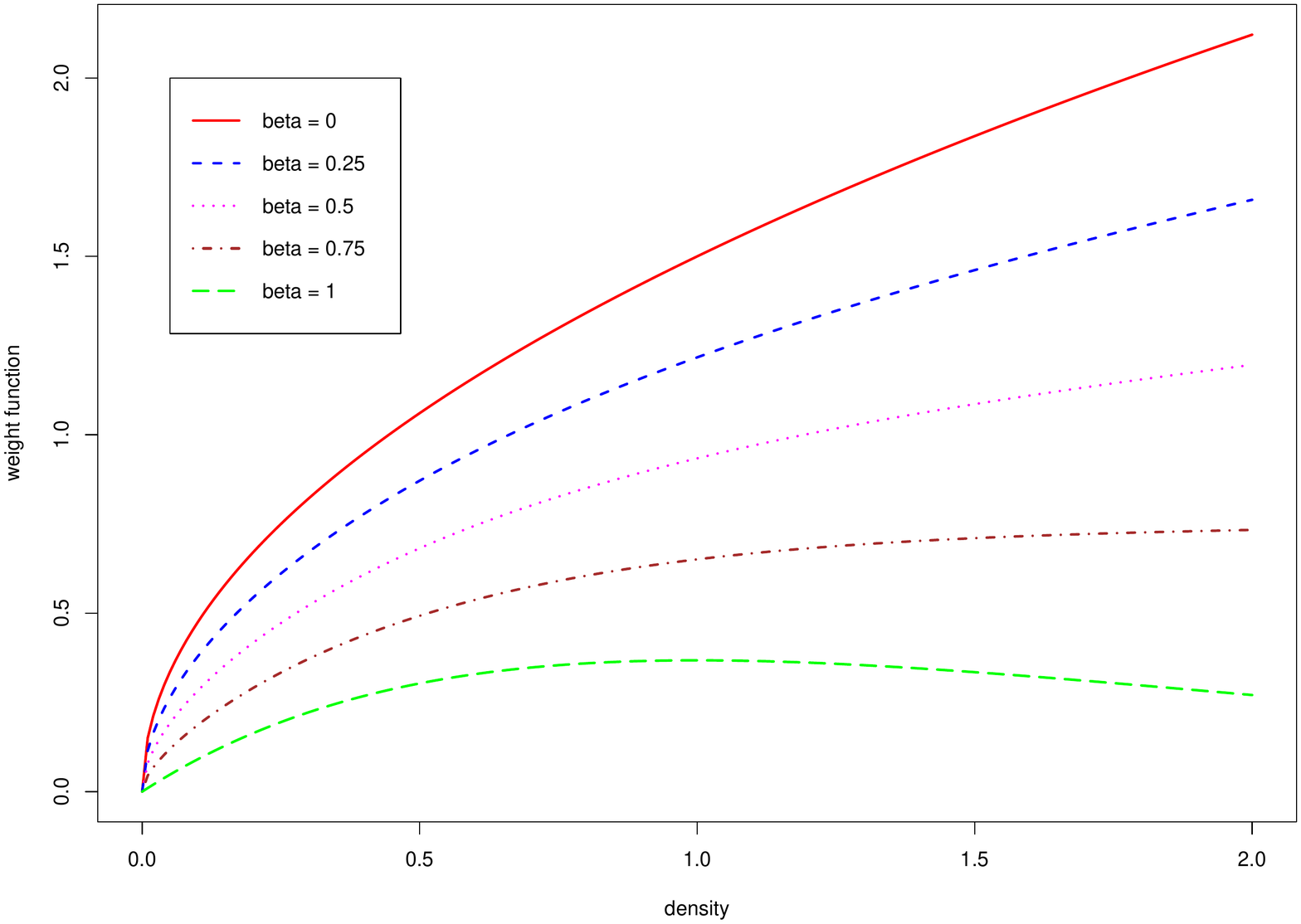}\vspace{-.2cm}\\ \vspace{-.1cm}
			(c) &  (d)
		\end{tabular}
		\caption{The weight function of the MEPDE for different values of $\beta$ when (a) $\alpha=1$ and $\gamma=1$, (b) $\alpha=-1$ and $\gamma=1$, (c) $\alpha=1$ and $\gamma=0.5$, (d) $\alpha=-1$ and $\gamma=0.5$.}
		\label{fig:weights}
	\end{figure}

	A comparison of Equations (\ref{wx}), \eqref{B} and (\ref{tx}) show that the weight function of the estimating equation in case of the EPD has the form 
	$$w_\theta(x) = w(f_\theta(x)) =\beta f_\theta(x) \exp(\alpha f_\theta(x)) 
	+ (1-\beta)(\gamma+1) f_\theta^\gamma(x) .$$ 
	In Figure \ref{fig:weights} we give a description of some weight functions for different triplet combinations, with particular emphasis on what the variation in the parameters $\alpha$ and $\beta$ do to the estimation procedure when the value of $\gamma$ is kept fixed. From the figure it may be noted that between the variety of cases considered, downweighting patterns of many different types are observed.  In particular, in comparison to the weight function of the MDPDE (which corresponds to $\beta = 0$, irrespective of the value of $\alpha$), all different kinds of variations are observed. One set of procedures apply greater downweighting for less probable observations while increasing the weights of the others. On the other hand, others exhibit a greater smoothing effect leading to more uniform weight functions. On the whole, there is a medley of possibilities, from which the experimenter can choose the optimal procedure in a given situation. 
	
	\section{Independent and Identically Distributed (IID) Models}
	In this section, we will consider the parametric set up of Section \ref{sec:Breg_div} where an independent and identically distributed sample $X_1, X_2, \cdots, X_n$ is available from the true distribution $G$, which is modelled by the parametric family $\mathscr{F}=\{F_{\theta} : \theta \in \Theta \subset{\mathbb{R}}^p \}$. When the true distribution belongs to the model, so that $G = F_{\theta} $ for some $\theta \in \Theta $, the formulae for $J$, $K$ and $\xi$  defined in Equation (\ref{J_K_psi}) simplify to
	
	\begin{equation}
	\label{aysmptotic variance}
	\begin{aligned}
	J(\theta) &= \beta \int_x u_{\theta}(x) u_{\theta} ^ T (x) f_{\theta}^{2}(x) \exp(\alpha f_{\theta}(x)) dx + (1-\beta)(\gamma+1) \int_x u_{\theta}(x) u_{\theta} ^T (x) f_{\theta}^{\gamma+1}(x) dx, \\ 
	K(\theta) &= \int_x  u_{\theta}(x)u_{\theta}^{T}(x) \big\{\beta f_{\theta}(x) \exp(\alpha f_{\theta}(x)) + (1-\beta)(\gamma+1) f_{\theta}^{\gamma}(x)\big\}^{2}f_{\theta}(x)  dx, \\
	\xi(\theta) &= \int_x u_{\theta}(x) \big\{\beta f_{\theta}(x) \exp(\alpha f_{\theta}(x)) + (1-\beta)(\gamma+1) f_{\theta}^{\gamma}(x) \big\}  f_{\theta}(x)  dx.
	\end{aligned}
	\end{equation}
	
	When $\beta = 0 $ and $\gamma \downarrow 0$, $J(\theta)$ and $K(\theta)$ coincide with $I(\theta)$, the Fisher information matrix, and the asysmptotic variance $J^{-1} K J^{-1}$ coincides with $I^{-1}(\theta)$, the inverse of the Fisher information.  The choice $\beta = 0$ leads to the variance estimates of MDPDE($\gamma$), while the choice $\beta = 1$ leads to the variance estimates of MBEDE($\alpha$), the minimum BED estimator for tuning parameter $\alpha$. 
	
	\subsection{Selecting the Optimal Procedure}
	
	What we have done so far in our development is that we have created a sophisticated Br{\`e}gman function which is a convex combination of the Br{\`e}gman functions of the DPD and the BED families, and described the related inference procedure. As the DPD is widely recognized as the current standard in density-based minimum distance inference based on divergences of the Br{\`e}gman type, our main motivation is to show that our exploration allows us, in any given real situation, to select a procedure, which provides a better control in comparison to the procedures restricted to the DPD class.  
	
	Be it in the case of parametric estimation based on the density power divergence or the exponential-polynomial divergence, these estimation schemes allow millions of choices as they are indexed by one or more tuning parameters that are allowed to vary over some continuous range. The collection of procedures involves all different kinds of methods, ranging from the most efficient to highly robust ones. Yet, in any particular real data problem, the experimenter has to provide a single, most appropriate choice for the tuning parameter for the specific data at hand, without knowing the amount of anomaly that is involved in the data under consideration. In an intuitive sense it is clear that such choices should be data-based.   
	
	\subsection{The Current State-of-the-art}
	
	In robust statistical inference, which depends on one or more tuning parameters, a perennial problem is to choose the tuning parameter(s) appropriately when it has to be applied to a given set of numerical data. Such tuning parameters inevitably control the trade off between efficiency and robustness, and depending on what is needed and to what extent in a particular situation, the tuning parameter must strike a balance between these two conflicting requirements. 
	
	With the success of the DPD as a method of choice in robust statistical inference, several methods for the selection of the ``optimal''  DPD tuning parameter has been proposed in the literature. The basic idea is the construction of an empirical measure of mean square error (or some other similar objective) as a function of the tuning parameter, which can then be minimized over the latter; this generates a minimum mean square error criterion for the selection of the tuning parameter. A few variations of this technique has been tried out in the literature. Here we will follow the approach considered in \cite{warwick2005choosing}.  
	
	In the above approach, we will evaluate the performance of the estimator through its summed mean square error around $\theta^*$, which may be expressed, asymptotically, as 
	\begin{equation}\label{summed_mse}
	E\left((\hat{\theta}_n - \theta^*)^T(\hat{\theta}_n - \theta^*)
	\right) = n^{-1}{\rm tr}\left(J^{-1}(\theta^g) K(\theta^g) J^{-1}(\theta^g)\right)+ (\theta^g -\theta^*)^T(\theta^g-\theta^*),
	\end{equation}
	where $\hat{\theta}_n$, the MEPDE, is a function of triplet of tuning parameters $(\alpha, \beta, \gamma)$, $J$ and $K$ are as in Equation (\ref{J_K_psi}), and ${\rm tr}(\cdot)$ represents the trace of a matrix. Such a formulation may be meaningful, for example, when the data are generated by a mixture having $f_{\theta^*}$ as the dominant component, and $\theta^*$ is our target parameter; see the discussion in \cite{warwick2005choosing}.
	In practice we empirically estimate the quantity on the right hand side of Equation (\ref{summed_mse}) by replacing the true distribution $G$, wherever possible, by the empirical distribution $G_n$, $\theta^g$ with the MEPDE $\hat{\theta}_n$, and  $\theta^*$ by a suitable robust pilot estimator. In our calculations, following the suggestion of \cite{ghosh2015robust}, we will use the MDPDE at $\gamma = 0.5$ as the pilot estimator. This gives us an empirical mean square error as a function of the tuning parameters (and the pilot estimator), which can then be minimized over the tuning parameters to obtain their ``optimal'' estimates. 
	
	\subsection{Examples}
	
	In this section we will look at several well known real data examples, and demonstrate that suitable members of the MEPDE family provide useful robust fits to these data. All of these data sets have one or more large outliers so that robust procedures are meaningful in this context. 
	
	\begin{example}[(Telephone Fault Data):]
		We consider the data on telephone line faults presented and analyzed by \cite{welch1987rerandomizing} and \cite{simpson1989hellinger}. The data set is made up of the ordered differences between the inverse test rates and the inverse control rates in 14 matched pairs of areas. A normal model may otherwise work very well for these data, but the first observation is a huge outlier, and estimation by the method of maximum likelihood leads to a complete mess. The MLEs of $\mu$ and $\sigma$ under the normal model are 40.3571 and 311.332 respectively. For the outlier deleted data these estimates shift to 119.46 and 134.82, respectively, indicating that the single outlier suffices to completely destroy the inference based on maximum likelihood.  The MEPDEs of $\mu$ and $\sigma$ based on the optimal Warwick Jones tuning parameters are 122.205 and 136.962, corresponding to the triplet $\alpha = 0.98$, $\beta = 0.367$ and $\gamma = 0.146$. Note that this tuning parameter triplet is somewhat removed from the DPD family, which corresponds to $\beta =0$. 
	\end{example}
	
	\begin{example}[(Newcomb Data):]
		This is an old data set representing Newcomb's measurements on the velocity of light over a distance of 3721 meters and back \citep{stigler1977robust}. The main cluster of the data is again well modeled by a normal distribution, but two individual outliers hinder the estimation based on maximum likelihood. The MLEs of $\mu$ and $\sigma$ for the full data equal  26.2121 and 10.6636, respectively; but with the removal of the two outliers they shift to 27.750  and 5.044, with the estimate of $\sigma$ taking a huge drop. The MEPDEs of $\mu$ and $\sigma$ for the optimal Warwick Jones method are 27.6036 and  4.99074, respectively, corresponding to the triplet (0.996, 0.422, 0.297) for $(\alpha, \beta, \gamma)$; it is again somewhat removed from the DPD family. 
	\end{example}

	\begin{example}[(Darwin Data):]
		Charles Darwin had performed an experiment to determine whether cross-fertilized plants have higher growth rates compared to self fertilized plants \citep{spiegelhalter1985exact}. Pairs of
		Zea mays plants, one self and the other cross-fertilized, were planted in pots, and after a certain time interval the height of each plant was measured. The paired differences (cross-fertilized minus self fertilized) of 15 such pairs of plants were considered in this example. Once again a normal model appers to be suitable for these data, except for two large outliers in the left tail. For the full data, the MLEs of $\mu$ and $\sigma$ are  20.9333 and 36.4645, respectively, but for the outliers deleted data they become 33 and 20.8103, respectively. The optimal Warwick Jones method selects a member of the DPD family in this case with $\gamma = 0.5353$. The corresponding estimates are $\hat{\mu} = 29.8026$ and $\hat{\sigma} = 25.2416$.

	\end{example}

	\begin{example}[(Insulating Fluid data):]
		This example represents data that may be well fitted by an exponential model \citep{nelson1972graphical}. It involves tests regarding times to breakdown of an insulating fluid between electrodes recorded at seven different voltages. We consider the observations corresponding to voltage 34 kV. We are interested in estimating the mean parameter under the exponential model.  The data set has 19 observations, containing four large outliers and one massive outlier.
		The full data MLE of the mean parameter is 14.3589,
		whereas after deleting the five outliers, the outlier deleted MLE is 4.6457. The optimum MEPDE, on the other hand, equals 8.1599, and corresponds to the triplet
		$(-33.0234, 1, 0.5878)$. In this case it may be seen that the optimal solution corresponds to $\beta = 1$, and therefore belongs to the BED family with no contribution from the DPD part. 
	\end{example}

	\section{Independent Non-homogeneous Observations}
	
	In real life problems we hardly encounter identically distributed  data. In parametric estimation we often deal with the data which is not identical. In this section we obtain general method of robust estimation for non-homogeneous  data. 
	\subsection{Introduction}
	In the previous sections, we assumed that the data are independent as well as homogeneous. Now we relax the condition of homogeneity (identical distribution) and obtain the estimation procedure to be used in such cases. 
	We consider the data $Y_1,Y_2, \cdots,Y_n$, where $Y_i$s are independent but each with a different density $g_i$. Our aim is to model $g_i$ by a family of distributions $\mathscr{F}_{i,\theta} = \{f_i(. {;} \theta) : \theta \in \Theta \subset \mathbb{R}^p\} $ for some $\theta$ for all $i = 1, 2, \cdots,n$, where $\theta$ is a parameter of interest. Thus although the $Y_i$s are not identically distributed, their distributions are based on a common parameter. Let us consider the Br{\`e}gman divergence defined in Equation \eqref{eqn:Bregman density form}. The MEPDE of $\theta$ is obtained by minimizing the empirical objective function
	\begin{equation}
	\frac{1}{n}\sum_{1=1}^{n}D_B(\hat{g}_{i},f_i(.;\theta)),
	\end{equation}
	over $\theta \in \Theta$, where $\hat{g}_i$ is an estimate of density $g_i$. Following Equation \eqref{gen_Bre_est}, it is sufficient to minimize
	\begin{align}
	H_n(\theta)
	= \frac{1}{n} \sum_{i=1}^{n}\Bigg[\int_y \Big\{B'(f_{i}(y,\theta))f_{i}(y,\theta) - B(f_{i}(y,\theta))\Big\} dy -B'(f_{i}(Y_{i},\theta))\Bigg] 
	=\frac{1}{n}\sum_{i=1}^{n}V_{i}(Y_{i},\theta),
	\label{reg_v}
	\end{align}
	where $V_{i}(.,\theta)$ is the term within the square brackets in Equation \eqref{reg_v}. It leads to the following estimating equation
	\begin{equation}
	\sum_{i=1}^{n} \bigg[ u_{i}(Y_{i};\theta)B''(f_{i}(Y_{i};\theta))f_{i}(Y_{i};\theta)-\int_y  u_{i}(y;\theta)B''(f_{i}(y;\theta)f_{i}^{2}(y;\theta) dy\bigg] =0,
	\label{est_eqn_non_homo}
	\end{equation}
	where $u_{i}(y;\theta) = \nabla_{\theta} \log(f_{i}(y;\theta))$. It can be viewed as a weighted likelihood estimating equation similar to Equation  \eqref{wt_lik}. In particular, by taking the $B$ function as given in Equation \eqref{B},  the estimating equation for the MEPDE is given by
	\begin{equation}
	\sum_{i=1}^n \left[u_i(Y_i;\theta)w(f_i(Y_i;\theta)) -\int_y u_i(y;\theta)w(f_i(y;\theta))f_i(y;\theta) dy\right]=0,
	\label{est_eqn_reg}
	\end{equation} 	
	where the weight function $w(t)= \beta t \exp(\alpha t) + (1-\beta)(1+\gamma) t^{\gamma}$. \cite{ghosh2013robust} derived the asymptotic distribution of the MDPDE in this setup. We will now generalize it for the EPD measure.

	\subsection{Asymptotic Properties} \label{sec:asymp_non_hom}
	Let us define  a $p\times p$ matrix $J^{(i)}$ whose $(k,l)$-th element is given by 
	\begin{equation}
	J^{(i)}_{kl} = E_{g_{i}}(\nabla_{kl}V_{i}(Y;\theta)), \mbox{ for } i = 1, 2, \cdots, n,
	\label{J_mat_Non_hom}
	\end{equation}
	where $\nabla_{kl}$ represents the partial derivative with respect to the $k$ and $l$-th element of $\theta$. We also define
	\begin{equation}
	\Psi_{n}= \frac{1}{n} \sum_{i=1}^{n} J^{(i)},
	\ \ \ 
	\Omega_{n}= n^{-1} \sum_{i=1}^{n} {\rm{Var}}_{g_{i}}\big(\nabla V_{i}(Y_{i},\theta) \big).
	\label{Omega_non_homo}
	\end{equation}
	Suppose $\theta^g$ is the best fitting parameter as defined in Section \ref{sec:m_est}. Following Equation \eqref{J_K_psi}, we can show that
	\begin{equation}
	\begin{aligned}
	J^{(i)}  =&  \beta \int_y f_{i}^2(y;\theta^g) \exp(\alpha f_{i}(y;\theta^g)) u_i(y;\theta^g) u_i^T(y;\theta^g) dy \\
	&+ (1-\beta)(\gamma+1) \int_y f_{i}^{\gamma+1}(y;\theta^g) u_i(y;\theta^g) u_i^T(y;\theta^g) dy\\ 
	&   + (1-\beta)(\gamma+1)\int_y \big( g_i(y)- f_{i}(y;\theta^g)\big) {\Big\{ I_i(y, \theta^g)-\gamma u_i(y;\theta^g) u_i^T(y;\theta^g)  \Big\} f_{i}^{\gamma}(y;\theta^g)} dy\\
	&  +\beta \int_y (g_i(y)-f_{i}(y;\theta^g))\Big\{ I_i(y, \theta^g) - u_i(y;\theta^g)u_i^T(y;\theta^g) \Big\}  f_{i}(y;\theta^g) \exp(\alpha f_{i}(y;\theta^g)) dy \\
	&  - \alpha \beta \int_y \big( g_i(y)-f_{i}(y;\theta^g)\big) f_{i}^{2}(y;\theta^g) \exp(\alpha f_{i}(y;\theta^g)) u_i(y;\theta^g) u_i^T(y;\theta^g) dy,	
	\end{aligned}
	\label{J_K_psi_nh}
	\end{equation}
	where $I_i(y, \theta^g)=-\nabla u_i(y;\theta^g)$ and	
	\begin{equation}
	\begin{aligned}
	\Omega_{n} = \frac{1}{n} \sum_{i=1}^n \Bigg[ &\int_y u_i(y;\theta^g)u_i^T(y;\theta^g)\Big\{\beta f_{i}(y;\theta^g)\exp(\alpha f_{i}(y;\theta^g)) \\
	&+ (1-\beta)(\gamma+1)f_{i}^{\gamma}(y;\theta^g)\Big\}^{2}g_i(y) dy - \xi_i\xi_i^{T} \Bigg],    
	\end{aligned}
	\label{omega_nh}
	\end{equation}
	\begin{equation}
	\begin{aligned}
	\xi_i =& \int_y u_i(y;\theta^g)\Big\{\beta f_{i}(y;\theta^g)\exp(\alpha f_{i}(y;\theta^g)) + (1-\beta)(\gamma+1)f_{i}^{\gamma}(y;\theta^g)\Big\} g_i(y) dy.
	\end{aligned}
	\label{xi_nh}
	\end{equation}

	\begin{theorem} \label{theorem2}
		Under the conditions (B1)--(B7) given in Appendix B the following results hold
		\begin{enumerate}
			\item[(a)] There exists a consistent sequence of solution $\hat{\theta}_{n}$  of Equation \eqref{est_eqn_non_homo}.
			\item[(b)] The asymptotic distribution of  $\Omega_{n}^{-1/2}\Psi_{n}[\sqrt{n}(\hat{\theta}_n-\theta_g)]$ is $p$-dimensional normal with mean (vector) 0 and covariance matrix $I_{p}$, the $p$-dimensional identity matrix.
		\end{enumerate}
	\end{theorem}

	\begin{remark} The proof of this theorem is similar to that of Theorem 3.1 of \cite{ghosh2013robust}. Theorem \ref{theorem1} is a special case of Theorem \ref{theorem2} if we assume an IID model, i.e. $f_{i}(\cdot; \theta) = f(\cdot; \theta)$, for all $i = 1, 2, \cdots,n$. The asymptotic distribution of the MDPDE derived by \cite{ghosh2013robust} also emerges as a special case of this theorem for $\beta = 0$. 
	
\end{remark}
	
	\subsection{\textbf{Linear Regression}}
	The theory proposed above can be readily applied to the case of  linear regression. 
	Consider the linear regression model
	\begin{equation}
	Y_i = x_i^T\eta + \epsilon_i, \ \ i=1,2, \cdots, n,
	\end{equation} 
	where the error $\epsilon_i$’s are IID errors having $N(0,\sigma^2)$ distributions. Here $x_i$'s are fixed design variables and $\eta=(\eta_1, \eta_2, \cdots, \eta_p)^T$ represents the regression
	coefficient. The parameter of our interest is $\theta = (\eta^T, \sigma^2)^T$. Note that $Y_i$'s are independent but not identically distributed random variables as $Y_i \sim f_i(.; \theta)$, where  $f_i(.; \theta)$ is $N(x_i^T\gamma,\sigma^2) $ distribution.  The score function for the normal model is given by
	\begin{equation}
	u_i(Y_i; \theta) = \left( \begin{array}{c} 
	\frac{(Y_i- x_i^T\eta)}{\sigma^2}x_i \\
	\frac{(Y_i - x_i^T\eta)^2-\sigma^2}{2\sigma^4}
	\end{array} \right).
	\end{equation}  
	So, the estimating equation \eqref{est_eqn_reg} simplifies as
	\begin{equation}
	\sum_{i=1}^{n} x_{ij} (Y_{i}-x^T_i \eta)\Big[\beta f_i(Y_i;\theta) \exp(\alpha f_i(Y_i;\theta)) + (1-\beta)(1+\gamma) f_i^{\gamma}(Y_i;\theta)\Big]
	=0, \ \  j=1,2,\cdots,p,     
	\end{equation}
	\begin{equation}
	\begin{split}
	\sum_{i=1}^{n} & \Big\{(Y_{i}-x^T_i \eta)^2-\sigma^2 \Big\} \Big[\beta f_i(Y_i;\theta) \exp(\alpha f_i(Y_i;\theta)) + (1-\beta)(1+\gamma) f_i^{\gamma}(Y_i;\theta)\Big]\\
	& =   
	\sum_{i=1}^{n} \int_y  \Big\{(y-x^T_i \eta)^2-\sigma^2\Big\} \Big[\beta f_i^2(y;\theta) \exp(\alpha f_i(y;\theta)) + (1-\beta)(1+\gamma) f_i^{\gamma+1}(y;\theta) \Big] dy.
	\end{split}
	\end{equation}
	To obtain the asymptotic distribution of the MEPDE, for simplicity, we assume that the true data generating
	density $g_i$  belongs to the model family of distributions, i.e., $g_i=f_i(.;\theta)$ for all $i=1,2,\cdots, n$, and $\theta = (\eta^T, \sigma^2)^T$ is the true value of the parameter. It simplifies $J^{(i)}$ in Equation \eqref{J_K_psi_nh} 	to	
	\begin{equation}
	\begin{aligned}
	J^{(i)}  =&  \beta \int_y f_{i}^2(y;\theta) \exp(\alpha f_{i}(y;\theta)) u_i(y;\theta) u_i^T(y;\theta) dy \\
	&+ (1-\beta)(\gamma+1) \int_y f_{i}^{\gamma+1}(y;\theta) u_i(y;\theta) u_i^T(y;\theta) dy.
	\label{reg_J}
	\end{aligned}
	\end{equation}
	It gives
	\begin{equation}
	\begin{aligned}
	\Psi_n= &\begin{bmatrix} 
	\frac{\omega_1}{n} X^T X & 0 \\
	0 & \omega_2  \\		
	\end{bmatrix},
	\label{reg_J}
	\end{aligned}
	\end{equation}
	where $X^T = (x_1, x_2, \cdots, x_n)_{p\times n}$ is the transpose of the design matrix and 
	\begin{equation}
	\begin{split}
	\omega_1 = &
	\int_y   \frac{y^2}{\sigma^4}  \Big[\beta \phi^2(y;\sigma) \exp(\alpha \phi(y;\sigma)) + (1-\beta)(1+\gamma) \phi^{\gamma+1}(y;\sigma)\Big] dy,\\
	\omega_2 = & 
	\int_y   \frac{(y^2 -\sigma^2)^2}{4\sigma^8}  \Big[\beta \phi^2(y;\sigma) \exp(\alpha \phi(y;\theta)) + (1-\beta)(1+\gamma) \phi^{\gamma+1}(y;\sigma)\Big] dy,
	\end{split}		
	\end{equation}
	with $\phi(\cdot, \sigma)$ being the probability density function of $N(0,\sigma^2)$. 
	Similarly, $\Omega_n$ in Equation \eqref{omega_nh} simplifies to 
	\begin{equation}
	\begin{aligned}
	\Omega_{n} =& \frac{1}{n} \sum_{i=1}^n \Bigg[ \int_y u_i(y;\theta)u_i^T(y;\theta)\Big\{\beta f_{i}(y;\theta)\exp(\alpha f_{i}(y;\theta)) \\
	&+ (1-\beta)(\gamma+1)f_{i}^{\gamma}(y;\theta)\Big\}^{2}f_{i}(y;\theta) dy - \xi_i\xi_i^{T} \Bigg], \\
	= &\begin{bmatrix} 
	\frac{\omega_3}{n} X^TX & 0 \\
	0 & \omega_4  \\		
	\end{bmatrix},
	\label{reg_j_omega}
	\end{aligned}
	\end{equation}
	where
	\begin{equation}
	\begin{split}
	\omega_3 = &
	\int_y   \frac{y^2}{\sigma^2}  \Big\{\beta \phi(y;\sigma) \exp(\alpha \phi(y;\sigma)) + (1-\beta)(1+\gamma) \phi^{\gamma}(y;\sigma)\Big\}^2 \phi(y;\sigma) dy,\\
	\omega_4 = & 
	\int_y   \frac{(y^2 -\sigma^2)^2}{4\sigma^8}  \Big\{\beta \phi(y;\sigma) \exp(\alpha \phi(y;\sigma)) + (1-\beta)(1+\gamma) \phi^{\gamma}(y;\sigma)\Big\}^2 \phi(y;\sigma) dy\\
	& 
	- \Bigg[\int_y   \frac{y^2 -\sigma^2}{2\sigma^4}  \Big\{\beta \phi(y;\sigma) \exp(\alpha \phi(y;\sigma)) + (1-\beta)(1+\gamma) \phi^{\gamma}(y;\sigma)\Big\} \phi(y;\sigma) dy\Bigg]^2.
	\end{split}		
	\end{equation}		
	
	Under the conditions (B1)--(B7) of Appendix B, we conclude from Theorem \ref{theorem2} that the MEPDE $\hat{\theta}_n$ is a consistent estimator of $\theta$. Moreover, the asymptotic distribution of $\sqrt{n}\Omega_n^{-1/2}\Psi_n(\hat{\theta}_n - \theta)$ is  multivariate normal with mean (vector) zero and covariance matrix $I_{p}$.
	
		\begin{figure}
		\begin{center}
			\includegraphics[scale=0.75]{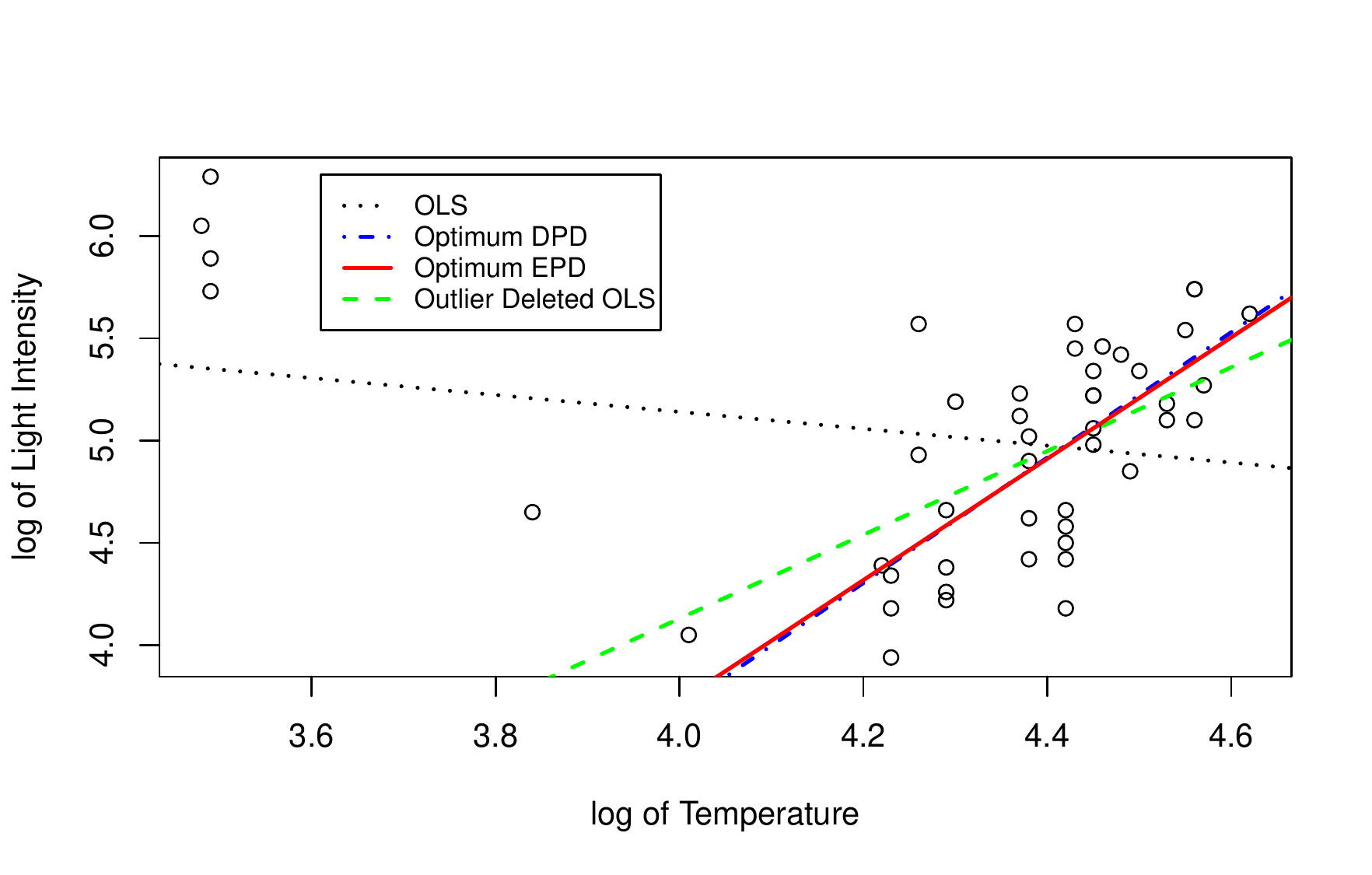}
			\caption{Plots of different regression lines for the Hertzsprung-Russell data of the star cluster.}
			\label{fig:star}
		\end{center}
	\end{figure}

	\subsection{Examples}
	We will give two examples to demonstrate the application of our proposed method in the independent  non-homogeneous data. These data sets are also analyzed by \cite{ghosh2013robust}.
	
	\begin{example}[(Hertzsprung-Russell data of the star cluster):] Our first data set contain 47 observations based on the Hertzsprung-Russell diagram of the star cluster CYG OB1 in the direction of Cygnus \citep{rousseeuw2005robust}.  We consider a simple linear regression model using the logarithm of the effective temperature at the surface of the star ($x$), and the logarithm of its light
		intensity ($y$). The scatter plot in Figure \ref{fig:star} shows that there are two groups of stars with four observations on the upper right corner  clearly separated from  others. In astronomy, those four stars are known as giants. The values of different regression estimates are given on Table \ref{tab:star}, and the fitted regression lines are added on Figure \ref{fig:star}. Due to four large outliers, the ordinary least squares (OLS) method completely fails to fit the data set. But the outliers deleted OLS gives a good fit for the rest of the 43 observations. Both the optimum DPD and EPD fits based on the Warwick Jones method are also close to that line. Here the optimum EPD corresponds to the triplet ($-4.8715, 0.9897, 0.7558$) for ($\alpha, \beta, \gamma$), whereas the optimum DPD parameter is $\gamma=0.75$. So, the optimum MEPDE lies well outside the DPD family. Also note that the estimate of $\sigma$ is much sharper in case of the MEPDE compared to the MDPDE, indicating that the former does much better than the latter in downweighting the outliers. 
		
	\end{example}
\begin{table}
	\begin{center}
		\begin{tabular}{l|rrr}
			\hline
			Methods & $\hat{\eta}_0$ & $\hat{\eta}_1$ & $\hat{\sigma}^2$\\
			 \hline 
			OLS & 6.7935 & $-0.4133$ &0.3188 \\
			Optimum DPD & $-8.5570$ & 3.0622 & 0.1616\\
			Optimum EPD & $-8.1389$ & 2.9660 & 0.1035\\
			Outlier Deleted OLS & $-4.0565$ & 2.0467 & 0.1647\\
			\hline                         
		\end{tabular}
		\caption{Different regression estimates for the Hertzsprung-Russell data of the star cluster.}
		\label{tab:star}
	\end{center}
\end{table}

\begin{example}[(Belgium telephone call data):] We consider a real data set from the Belgian Statistical Survey published by the 	Ministry of Economy of Belgium; it is also available in \cite{rousseeuw2005robust}. It contains the total number (in tens of millions) of international phone calls made in a year from 1950 to 1973. There is a heavy contamination in the vertical axis due to the use of a different
	recording system during 1964 to 1969. The years 1963 and 1970 are also partially affected for this reason. Figure \ref{fig:telephone} and Table \ref{tab:telephone} contain the different regression estimates for this data set. It is clear that the OLS fit is very poor, but all other estimates give excellent fits to the rest of the observations. Although, the optimum EPD regression line based on the Warwick Jones method almost coincides with the optimum DPD fit, the MEPDE does not belong to the DPD family. The optimum EPD corresponds to the triplet ($-4.2416,0.0543, 0.3205$) for ($\alpha, \beta, \gamma$), whereas the optimum DPD parameter is $\gamma=0.631$. Once again the MEPDE produces a sharper value of the estimate of $\sigma$ compared to the MDPDE. 

\begin{figure}
	\begin{center}
		\includegraphics[scale=0.75]{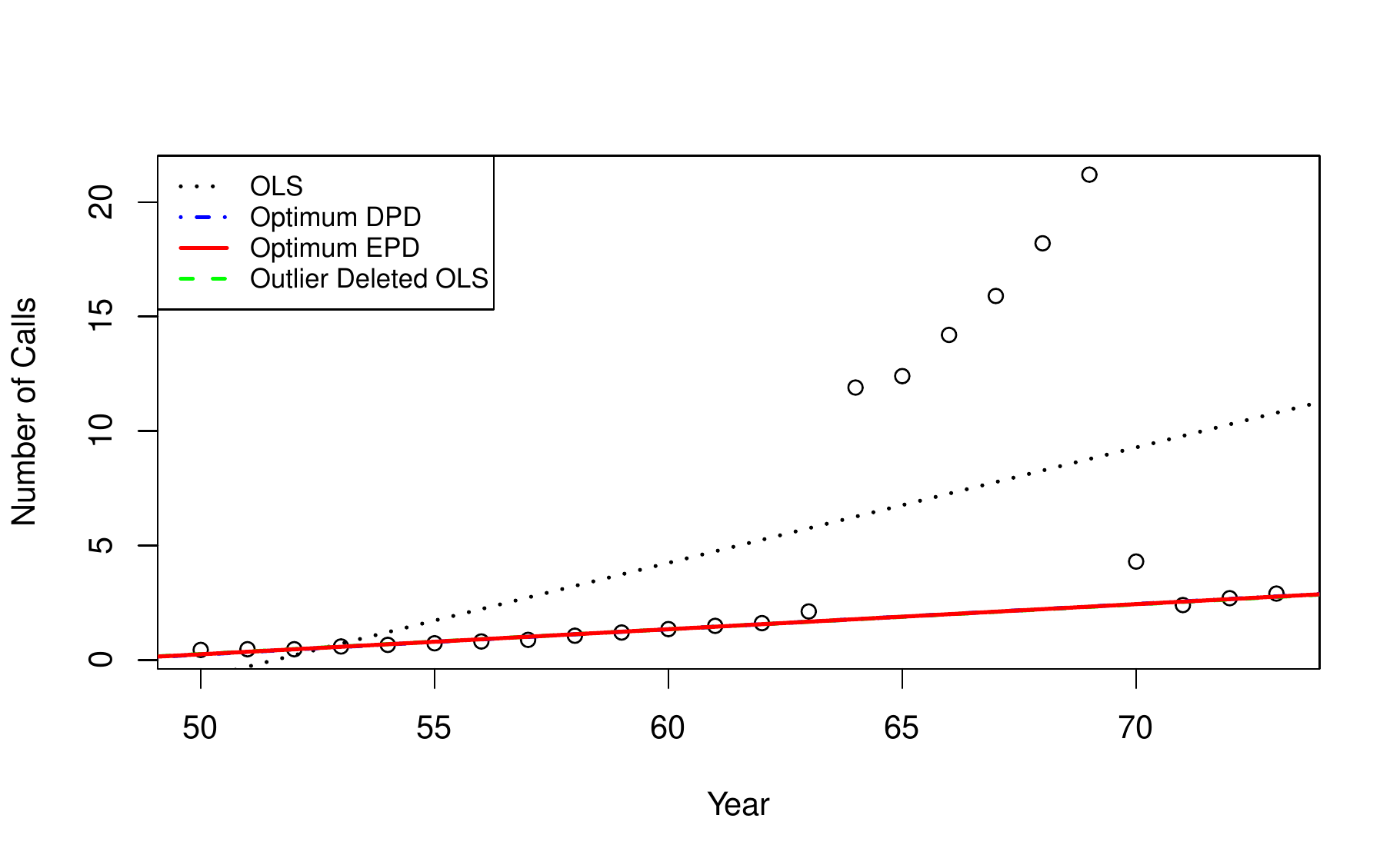}
		\caption{Plots of different regression lines for the Belgium telephone call data.}
		\label{fig:telephone}
	\end{center}
\end{figure}

\begin{table}
	\begin{center}
		\begin{tabular}{l|rrr}
			\hline
			Methods & $\hat{\eta}_0$ & $\hat{\eta}_1$ & $\hat{\sigma}^2$\\
			\hline 
			OLS & $-26.006$ &  0.5041 & 31.6107  \\
			Optimum DPD & $-5.2811$ & 0.1104 & 0.01336\\
			Optimum EPD & $-5.2278$ & 0.1095 & 0.0123\\
			Outlier Deleted OLS & $-5.1645$ & 0.1085 & 0.0094\\
			\hline                         
		\end{tabular}
		\caption{Different regression estimates for the Belgium telephone call data.}
		\label{tab:telephone}
	\end{center}
\end{table}
\end{example}
	
	\section{Concluding Remarks}
	
	Density-based minimum distance procedures have become popular in recent times because of their ability to combine high asymptotic efficiency with strong robustness properties. In particular the methods based on the Br{\`e}gman divergence have the major advantage that they do not involve any intermediate non-parametric smoothing component.  The class of DPD family, which has proved to be a popular and useful tool in this area, represents a class of procedures ranging from highly efficient to strongly robust. In this paper we have developed a more refined class of divergences which subsumes the DPD family providing new options which can lead to better compromises between robustness and efficiency.  
	
	In this paper we have demonstrated the above through IID data models as well as INH models. The results show that in most cases the optimal solution is outside the DPD family. These can, however, be extended to many other data structures where the EPD can be useful. For example, this technique can be used to find the best tuning parameter in estimation with right censored survival data, and testing of hypothesis problems, issues that we want to deal with in the future. 
	
	We also hope to use a recently developed refinement of the Warwick and Jones approach, present in \cite{basak2020optimal}, for the ``optimal'' tuning parameter selection problem, which might further enhance the results of our method. 
	
	\section*{Conflict of interest}
	
On behalf of all authors, the corresponding author states that there is no conflict of interest.
	
	\bibliographystyle{apalike}
	\bibliography{bibliography}

	\begin{appendices}
		\section{Conditions for Theorem \ref{theorem1}}
		For any given values of parameters $(\alpha,\beta , \gamma)$, we assume  the following conditions as an extension of conditions  given in \cite{basu2011statistical} for MDPDE($\alpha$)
		\begin{enumerate}
			\item[(A1)] The distributions $F_{\theta}$ of $X$ have a common support, such that the set $\chi= \{ x: f_{\theta}(x)>0 \} $ is independent of $\theta$. The true distribution $G$ is also supported on $\chi$ where $g$ is positive. 
			
			\item [(A2)] There is an open subset $\omega$ of the parameter space $\Omega $ containing the best fitting parameter $\theta^g$  such that for almost all $x \in \chi $ and all $ \theta \in \Theta $, the density $f_{\theta}(x) $ is three times differentiable with respect to $\theta$ and the third partial derivatives are continuous with respect to $\theta $.
			 
			\item[(A3)] For the $B$ function given in Equation \eqref{B}, the integrals $\int_x \{f_{\theta}(x)B'(f_{\theta}(x)) - B(f_{\theta}(x))\}dx $ and $\int_x B'(f_{\theta}(x))g(x) dx $ can be differentiated three times with respect to $\theta$ and the derivatives can be taken under the integral sign.
			\item[(A4)] For  $B$  in Equation \eqref{B} and
			\begin{equation}
			V_{\theta}(X) = \int_x \Big\{ B'(f_{\theta}(x))f_{\theta}(x) - B(f_{\theta}(x))\Big\}dx - B'(f_{\theta}(x))  ,
			\end{equation}
			the   $p\times p$ matrix defined by 
			$	J_{kl}(\theta) = E_{g} ( \nabla_{kl}V_{\theta}(X) ) $			
			is positive definite, where $E_{g}$ represents the expectation under the density $g$. When $g$ is in the model, then
			$J_{kl}(\theta^{g}) = J_{kl}(\theta^{g})$, where $J(\theta)$ is as defined in \eqref{J_K_psi}.
			\item[(A5)]
			There exists a function  $ M_{jkl}(x) $ such that
			$ |{ \nabla_{jkl}}V_{\theta}(X) | \leq M_{jkl}(X)$    for all $\theta\in \omega,$  and $E_{g}[M_{jkl}(X)] = m_{jkl} < \infty .$
		\end{enumerate}
		\section{Conditions for Theorem \ref{theorem2}}
		
		The following assumptions are required to establish the
		asymptotic properties of the MEPDE for the non-homogeneous case. These are analogous to the assumptions given in \cite{ghosh2013robust} for the DPD family.
		
		\begin{enumerate}
			\item[(B1)] 			
			The support $\chi = \{y : f_i(y; \theta) > 0 \}$ is independent of $i$ and $\theta$ for all $i=1,2,\cdots, n$, and  the true distribution of $G_{i}$ is also supported on $\chi$ for all $i$.
			\item [(B2)] There is an open subset $\omega$ of the parameter space $\Omega $ containing the best fitting parameter $\theta^g$  such that for almost all $x \in \chi $ and all $ \theta \in \Theta $, the densities $f_{i}(y; \theta), \ i=1,2,\cdots, n$, are three times differentiable with respect to $\theta$ and the third partial derivatives are continuous with respect to $\theta $.
			
			\item[(B3)] Consider the $B$ function given in Equation \eqref{B}. For each $i=1,2,\dots,n$, the integrals $\int_y\big[B'(f_{i}(y;\theta))f_{i}(y;\theta)-B(f_{i}(y;\theta)\big]dy$ and $\int_y B'(f_{i}(y;\theta))g_{i}(y)dy $ can be differentiated thrice with respect to $\theta$ and derivatives can be taken under integral sign.
			\item[(B4)] For each $i=1,2,\dots, n$, the matrix $J^{(i)}$, defined in Section \ref{sec:asymp_non_hom}, is positive definite and 
			\begin{equation}
				\lambda_{0}= \inf_{n} \ [\mbox{min eigenvalue of } \Psi_{n}] > 0.
			\end{equation}
			
			\item[(B5)]   There exists a function $M_{jkl}^{(i)}(Y)$ such that 
			\begin{equation} |\nabla_{jkl} V_{i}(Y;\theta) |\leq M^{(i)}_{jkl}(Y) ,
			\end{equation} 
			where $V_{i}(\cdot;\theta)$ is defined in Equation \eqref{reg_v} and
			\begin{equation}
				\frac{1}{n} \sum_{i=1}^{n} E_{g_{i}}[M_{jkl}^{(i)}(Y)]=O(1)  \; \mbox{ for all } j,k,l.
			\end{equation} 
			
			\item[(B6)] For all $j$ and $k$, we have				
			\begin{align}
			\begin{split}
			\centering
			\lim_{N\rightarrow \infty} \sup_{n} \bigg(\frac{1}{n}E_{g_i}\big[|\nabla_{jk}V_{i}(Y;\theta)| I(|\nabla_{jk}V_{i}(Y;\theta)|>N)]\big]\bigg)&=0, \\
			\lim_{N\rightarrow \infty} \sup_{n} \bigg(\frac{1}{n}E_{g_i}\big[|\nabla_{jk}V_{i}(Y;\theta)-E_{g_{i}}(\nabla_{jk}((Y;\theta))| \\
			\times |I(|\nabla_{jk}V_{i}(Y;\theta)-E_{g_{i}}(\nabla_{jk}((Y;\theta))|>N)\big]\bigg)&=0 ,
			\end{split}
			\end{align}
			where $I(B)$ denotes the indicator variable of the event $B$.
			\item[(B7)] For all $\epsilon>0$, we have 
			\begin{equation}
			\lim_{n\rightarrow \infty}  \bigg\{ \frac{1}{n}\sum_{i=1}^{n} E_{g_{i}}\big[||\Omega_{n}^{-1/2} \nabla V_{i}(Y;\theta)||^{2} I(||\Omega_{n}^{-1/2} \nabla V_{i}(Y;\theta)||>\epsilon\sqrt{n})\big]\bigg\}=0.
			\end{equation}
		\end{enumerate}

	\end{appendices}
\end{document}